\documentclass[10pt, conference, letterpaper]{IEEEtran}

\usepackage{listings}
\usepackage{xcolor}

\definecolor{codegreen}{rgb}{0,0.6,0}
\definecolor{codegray}{rgb}{0.5,0.5,0.5}
\definecolor{codepurple}{rgb}{0.58,0,0.82}
\definecolor{backcolour}{rgb}{0.95,0.95,0.92}

\lstdefinestyle{mystyle}{
    backgroundcolor=\color{backcolour},   
    commentstyle=\color{codegreen},
    keywordstyle=\color{magenta},
    numberstyle=\tiny\color{codegray},
    stringstyle=\color{codepurple},
    basicstyle=\ttfamily\footnotesize,
    breakatwhitespace=false,         
    breaklines=true,                 
    captionpos=b,                    
    keepspaces=true,                 
    numbers=left,                    
    numbersep=5pt,                  
    showspaces=false,                
    showstringspaces=false,
    showtabs=false,                  
    tabsize=2
}

\usepackage[noadjust]{cite}
\usepackage{amsmath,amssymb,amsfonts}
\usepackage{algorithmic}
\usepackage{graphicx}
\usepackage{textcomp}
\usepackage{xcolor}
\usepackage{balance}
\usepackage{multirow}
\def\BibTeX{{\rm B\kern-.05em{\sc i\kern-.025em b}\kern-.08em
    T\kern-.1667em\lower.7ex\hbox{E}\kern-.125emX}}

\usepackage[some,bottom]{background}
\SetBgContents{\parbox{24cm}{\textbf{NOTE}: This paper has been accepted for publication at IEEE INFOCOM 2023. ©2023 IEEE. Personal use of this material is permitted. Permission from IEEE must be obtained for all other uses, in any current or future media, including reprinting/republishing this material for advertising or promotional purposes, creating new collective works, for resale or redistribution to servers or lists, or reuse of any copyrighted component of this work in other works.}}
\SetBgScale{0.6}
\SetBgOpacity{1}
\SetBgColor{gray}
\SetBgVshift{1.5cm}
\begin{document}

\title{ASTrack: Automatic Detection and Removal of Web Tracking Code with Minimal Functionality Loss}

\author{\IEEEauthorblockN{Ismael Castell-Uroz}
\IEEEauthorblockA{\textit{Universitat Polit\`ecnica de Catalunya} \\
Barcelona, Spain \\
ismael.castell@upc.edu}
\and
\IEEEauthorblockN{Kensuke Fukuda}
\IEEEauthorblockA{\textit{National Institute of Informatics} \\
Tokyo, Japan \\
kensuke@nii.ac.jp}
\and
\IEEEauthorblockN{Pere Barlet-Ros}
\IEEEauthorblockA{\textit{Universitat Polit\`ecnica de Catalunya} \\
Barcelona, Spain \\
pere.barlet@upc.edu}
}

\maketitle

\begin{abstract}
Recent advances in web technologies make it more difficult than ever to detect and block web tracking systems. In this work, we propose ASTrack, a novel approach to web tracking detection and removal. ASTrack uses an abstraction of the code structure based on Abstract Syntax Trees to selectively identify web tracking functionality shared across multiple web services. This new methodology allows us to: $(i)$ effectively detect web tracking code even when using evasion techniques (e.g., obfuscation, minification, or webpackaging); and $(ii)$ safely remove those portions of code related to tracking purposes without affecting the legitimate functionality of the website. Our evaluation with the top 10k most popular Internet domains shows that ASTrack can detect web tracking with high precision (98\%), while discovering about 50k tracking code pieces and more than 3,400 new tracking URLs not previously recognized by most popular privacy-preserving tools (e.g., uBlock Origin). Moreover, ASTrack achieved a 36\% reduction in functionality loss in comparison with the filter lists, one of the safest options available. Using a novel methodology that combines computer vision and manual inspection, we estimate that full functionality is preserved in more than 97\% of the websites.
\end{abstract}


\section{Introduction}
\BgThispage
During the last years, the research community has been very active looking for new ways to detect and block web tracking (e.g., \cite{acar_web_2014, englehardt_cookies_2015, li_trackadvisor_2015, metwalley_unsupervised_2015, nikiforakis_cookieless_2013, lerner_internet_2016, iqbal_fingerprinting_2021, castell-uroz_tracksign_2021}). Experts explored numerous online services, finding all kinds of new and exotic ways of exploiting protocols (\cite{acar_web_2014, englehardt_cookies_2015, li_trackadvisor_2015, metwalley_unsupervised_2015}) or abusing programming language APIs (\cite{nikiforakis_cookieless_2013, lerner_internet_2016, iqbal_fingerprinting_2021}) for user profiling purposes. Unfortunately, most of the proposed solutions are very complex and hard to deploy in a real browsing session, limiting their application to offline studies. Currently, the most popular solution to protect against web tracking is the use of content-filtering extensions in web browsers (e.g., uBlock Origin~\cite{hill_ublock_2020}, Adblock Plus~\cite{adblock_plus_adblock_2020}). These tools are based on \textit{filter lists}, manually curated pattern lists containing known tracking domains that are matched against the URLs visited in a browsing session.

However, content-filtering extensions as well as most of the previously proposed approaches suffer from several limitations: $(i)$ They require significant manual work to keep the filter lists up to date (e.g., new methods constantly emerge under different URLs); $(ii)$ URL-based detection and protection methods are easy to evade just by changing the hosting domain or dynamically modifying their URL parameters; $(iii)$ obfuscation, minimization, and webpackaging techniques automatically modify the internal website code, breaking many detection systems (e.g., \cite{wu_machine_2016, ikram_towards_2017}); and $(iv)$ current URL and resource-based blocking methods result in significant website functionality loss (\cite{krishnamurthy_measuring_2007, mazel_comparison_2019}). Some previous works (e.g., \cite{castell-uroz_tracksign_2021, iqbal_fingerprinting_2021, iqbal_adgraph_nodate, smith_sugarcoat_2021}) proposed partial solutions to some of these limitations, usually through the use of machine learning methods. However, these solutions present a trade-off, advancing in some aspects but giving up in others (see Section~\ref{background}). 

In this paper, we present ASTrack, a new method that addresses \textit{all} the limitations described above. Unlike previous proposals, ASTrack focuses on the code structure instead of the code itself. For this purpose, ASTrack uses an abstraction of the JavaScript code based on Abstract Syntax Trees (AST). An AST is simply a tree representation of the abstract syntactic structure of the source code, regardless of its particular contents (e.g., variable or function names). An AST can represent the entire code as well as the different functional portions of it (e.g., functions). Our proposal is based on the observation that most websites use common analytics or fingerprinting libraries to collect private information. Thus, when the same code structure (i.e., AST) is used across multiple domains, the AST becomes suspicious for performing tracking, especially if some of the domains were previously known to be tracking domains. By using a syntactic abstraction instead of the code itself, our system is more robust to common evasion techniques, such as minimization or obfuscation. Moreover, this abstraction allows us to selectively prune individual tracking ASTs (e.g., functions) while maintaining the rest of the (legitimate) code unmodified. This increase in blocking granularity compared to previous methods, which usually block URLs or complete resources, can better preserve website functionality and detect tracking code under different URLs or within different files.

Our evaluation of the top 10k most popular websites shows that ASTrack maintains a detection precision of more than 98\%. During our experiments, ASTrack found more than 3,400 new tracking URLs and automatically classified about 50k tracking code pieces, including obfuscated fragments that could not be easily detected with other techniques. Finally, we estimate that website functionality is preserved in approximately 97.7\% of websites.

In summary, the key contributions of this paper are:
\begin{enumerate}
    \item A \textbf{syntactic approach} to the detection of web tracking code that is highly adaptive and robust against minimization and obfuscation.
    \item A \textbf{high grade of detection granularity}, permitting selective code removal while maintaining the functionality of the website in most cases.
    \item A new \textbf{methodology to detect website functionality breakage} by means of computer vision techniques.
    \item An \textbf{evaluation of the tracking blocking performance as well as the website functionality loss} for the top 10k most popular websites in the Tranco List.
\end{enumerate}

The rest of the paper is organized as follows: Section \ref{background} presents an overview of web tracking detection systems and the limitations of existing methodologies. Section \ref{proposal} describes ASTrack, our new web tracking detection and removal proposal. Section \ref{evaluation} presents the evaluation of the web tracking detection and removal process. Finally, Section \ref{conclusions} concludes the paper and presents future work.

\BgThispage
\section{Background and Related work}
\label{background}
\subsection{Web tracking}
Traditional web tracking systems are usually \textit{stateful} technologies. They use different techniques to save an identifier inside the storage of the device browsing the web. The identifier will be read again every time the device accesses the same website service. The most common method are the infamous cookies, but there are more exotic approaches such as embedding identifiers in cached documents~\cite{implementing_web_tracking}, in the HTTP redirect cache~\cite{bursztein_tracking_nodate}, in the HTTP authentication cache~\cite{grossman_tracking_nodate}, or inside the HTML5 storage~\cite{ayenson_flash_2011}. However, new privacy regulations (e.g. \cite{gdpr, ccpa, standford_pipl_2021}) impose multiple restrictions on most of these systems. Moreover, many mainstream browsers have implemented countermeasures to this kind of web tracking. For instance, Safari now blocks all third-party cookies~\cite{safari_builtin}, and Firefox blocks third-party cookies from known trackers by default~\cite{firefox_builtin}. Chrome will also ban third-party cookies in the near future~\cite{google_cookies}.

\textit{Stateless} technologies, on the other hand, do not save information inside the device. They directly identify the user based on other measurable properties, such as the IP address or the device configuration exposed by the browser. Stateless technologies are also known as \textit{fingerprinting} methods. Most simple techniques look at numerous properties, such as the screen resolution~\cite{boda_user_2012}, the version of the browser~\cite{unger_shpf_2013} or the fonts installed in the system~\cite{fifield_fingerprinting_2015}, to combine them and create a unique identifier. However, the latest functionalities added to web technologies in the form of new JavaScript APIs~\cite{snyder_browser_2016} permit stateless web tracking algorithms to collect much more precise information. Rendering differences due to specific hardware and software combinations can be abused by means of those APIs to precisely identify the browser being used to explore the web \cite{mowery2012pixel, englehardt_online_2016, cao2017cross, starov_xhound_2017, zhu_eluding_2021}. This kind of web tracking is far more intrusive than the traditional cookies, as it is completely transparent to the user, and there is no easy way to control when, where, or by whom they are being used.

\subsection{Detection}

Most of the solutions proposed in the literature, such as the works from Wu et al.~\cite{wu_machine_2016} or Ikram et al.~\cite{ikram_towards_2017}, apply machine learning (ML) algorithms over the website code to find specific features identifying tracking methods. However, the underlying static analysis used is prone to fail under techniques such as obfuscation, that can modify those features.

Iqbal et al.~\cite{iqbal_fingerprinting_2021} present a \textit{browser fingerprinting} detection system using a combination of two independent machine learning algorithms, one trained with features extracted from a static analysis of the JavaScript code and the other from a dynamic analysis. The second ML complements the first one in case obfuscation or minimization techniques have been used in the code. They also present a way to reduce the breakage of website functionality by means of the replacement of the tracking code with mock versions of it. The system presents high accuracy in detecting browser fingerprinting and a functionality loss reduction of about 20\% whenever the mock functions can be used. However, the system only detects one kind of tracking, and the functionality breakage improvement system does not work on webpackaged files~\cite{noauthor_webpack_nodate}, a common practice on today's Internet. The breakage system is only evaluated on a population of 50 websites.

Our previous work at~\cite{castell-uroz_tracksign_2021} proposes a web tracking detection system based on a deterministic code partition algorithm and a tripartite network representation that allows us to propagate the probability of containing tracking for each piece of code. The system presents high accuracy in finding unknown web tracking. However, the static analysis nature of the system by using directly the code for identification makes it vulnerable to obfuscation or randomized content renaming techniques. Moreover, the partition algorithm makes it impossible to block tracking code without breaking website functionality.

\subsection{Mitigation}

Many of the detection systems presented above include some kind of solution to block the detected web tracking algorithms. However, those solutions are usually complex and difficult to deploy by the common user. In practice, the most commonly used mitigation systems are filter lists. These lists include a collection of patterns identifying URLs suspicious of performing tracking. There is no easy way to create the lists, and usually they are manually curated by the members of an online community. EasyList~\cite{open_source_easylist_2020} and EasyPrivacy~\cite{open_source_easyprivacy_2020} are both examples of such lists. Many popular content blockers, such as AdBlock Plus~\cite{adblock_plus_adblock_2020} and uBlock Origin~\cite{hill_ublock_2020} use those lists to block URLs during loading.

Smith et al. present in~\cite{smith_sugarcoat_2021} an alternative approach: tracking-free resource replacements. Some content blockers allow for resource replacements in real-time. Smith et al. automatically generate clean versions of some of the most popular tracking resources to be used as replacements. Similar to~\cite{iqbal_fingerprinting_2021}, they clean the code using a mock replacement for some of the API functions used to track the user. However, this approximation is not scalable, as the inspected scripts are selected manually, and many tracking systems work with dynamic custom files that are different for each website. Moreover, new tracking resources are created daily.

Some browsers also implement partial solutions to stateless tracking. In particular, the TOR browser and Firefox, by means of an experimental feature~\cite{firefox_fingerprinting}, automatically block some of the API functions commonly used for tracking~\cite{tor_fingerprinting}. Unfortunately, this approach breaks every website where the API is used for legitimate purposes.

\BgThispage
\section{ASTrack}
\label{proposal}

In this section, we introduce ASTrack, a new adaptive method to detect and selectively remove web tracking systems while minimizing the functionality loss associated with them. Our proposal is based on the observation that most websites share code and functionality, usually in the form of popular frameworks and useful libraries. For instance, JQuery libraries or social network interaction buttons are common to many different websites. Web tracking is no different. Most websites use common analytics or fingerprinting libraries (e.g., Google Analytics) to collect user information. It is rare to find websites with completely customized tracking libraries not present on other websites. Thus, web tracking functionality is also shared by many websites. Our proposal is to search for this shared functionality between multiple websites and automatically identify web tracking systems among them.

\subsection{Functionality identification}
As its name implies, ASTrack tracks shared Abstract Syntax Trees (AST) between different websites. An AST represents the code as a tree, whose nodes are its elements, and edges the relation between them. However, ASTrack does not look for the actual elements included in the AST but for the structure of the AST as a whole, looking only at the node types and their relations. The structure of the tree is independent of the names and values of the code, representing the functionality itself. Minimized or obfuscated ASTs share the same structure. Independent trees with exactly the same structure necessarily share the same functionality, despite the different results they may obtain depending on the input values. In summary, looking at AST's structure, we will find functionality shared by different web services.

\begin{figure}
  \centering
  \includegraphics[width=0.489\textwidth]{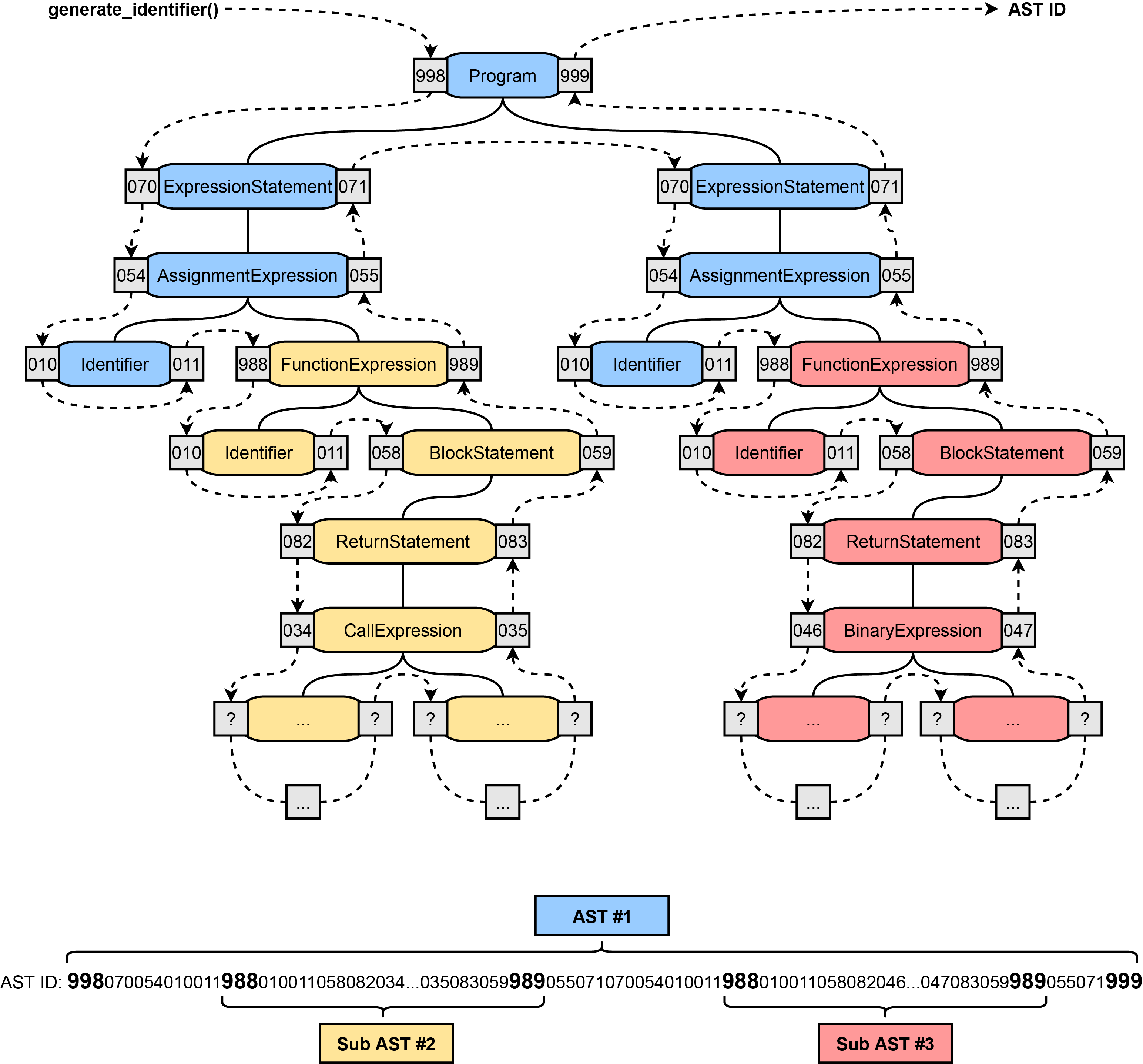}
  \caption{\textbf{AST identification example:} Integer values represent the aperture (left) and closure (right) labels of each node. The dotted arrow indicates the concatenation order to generate the ID. Looking for pairs of aperture and closure labels of function nodes, we find their corresponding identifiers.}
  \label{fig:ast_id}
  \vspace{-0.3cm}
\end{figure}

In order to define a representation of the AST structure, we decided to follow two principles: consistency and simplicity. In our case, consistency means that different codes with the same structure must always be represented equally. On the other hand, simplicity is a key factor in favoring a system that can be deployed in reality, where complex implementations can result in performance constraints. The selected representation corresponds to a simple \textbf{label chain}. The identifier label chain is created by traversing the AST, descending recursively into each branch, and concatenating integer identifiers for each node type found. Fig. \ref{fig:ast_id} shows a simple AST and the process to obtain its identifier for illustration purposes. Each node receives one aperture label (left integer) and one closure label (right integer), except operator nodes, which only have the aperture label. The dotted arrows indicate their concatenation order. When the algorithm enters a node, it concatenates its aperture label into the AST label chain. If the node has branches, it explores them recursively. Finally, if present, it also concatenates the closure label and finishes. The resultant label chain, composed of an ordered node identification chain, is the simplest representation that allows us to unambiguously identify the AST structure. However, not only the main AST identifies functionality, but each nested \textit{function} or \textit{method} also contains partial isolated functionality. Thus, we will generate identifiers for their branches as well. Fig.~\ref{fig:ast_id} shows three identifiers in different colors, one corresponding to the entire AST and two for internal nested function declarations. In order to find the identifier of the nested branches, we only need to look for the pair of labels that enclose them (988 and 989 in our example) within the main AST identifier. Finally, to fix the maximum length of the identifiers, a hash function is applied to each of their label chains. For purposes of simplification, during the rest of the paper we will refer to AST identifiers simply as \textit{AST}s.

\begin{figure*}
  \centering
  \includegraphics[width=\textwidth]{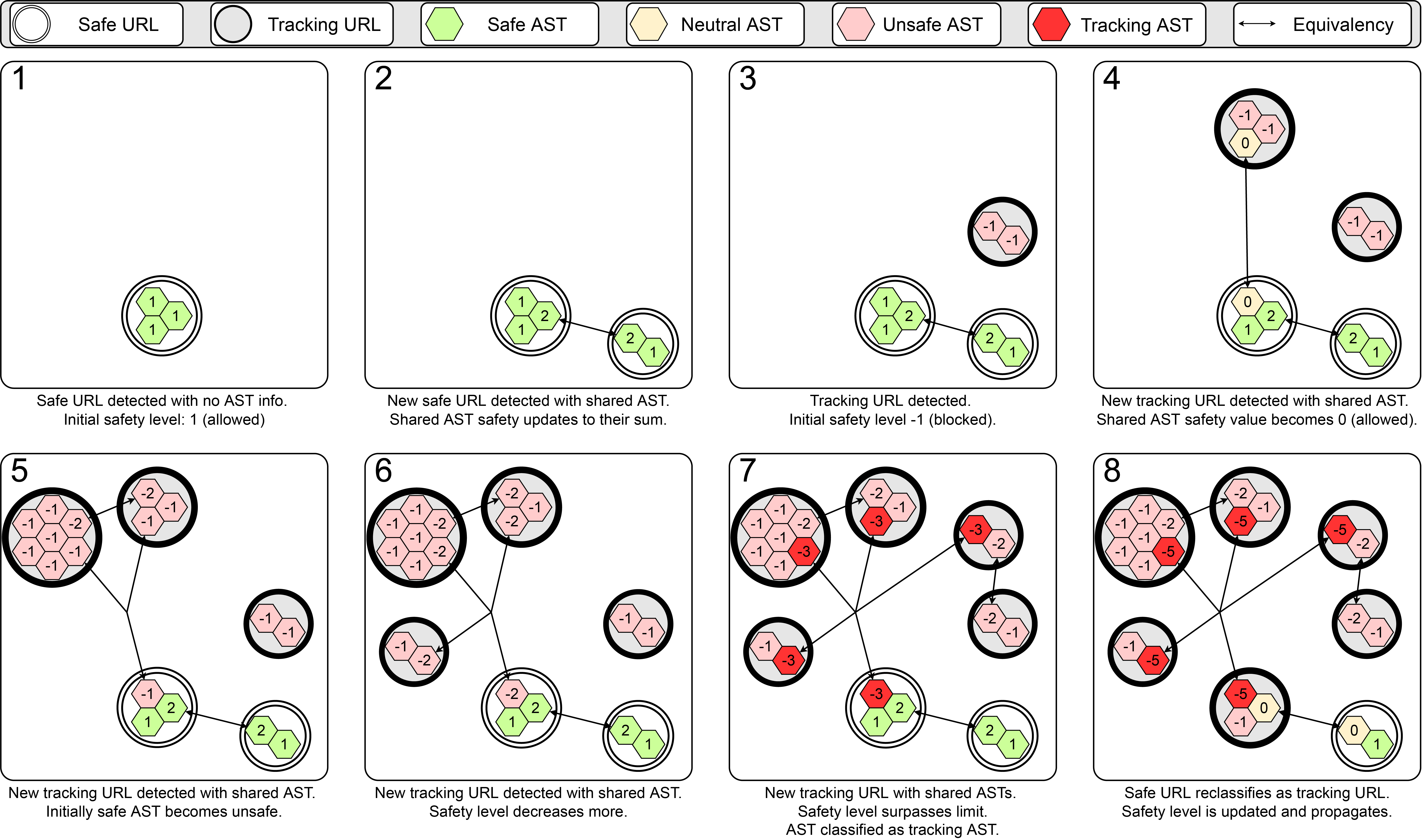}
  \caption{\textbf{ASTrack tracking detection and removal conceptual process:} By computing safety values for each AST loaded by URLs, ASTrack decides which code to remove or not. The safety is calculated by accounting for the number of already-known tracking URLs that include the AST. When the evidence is enough to consider the AST a tracking one, the URLs containing it are reclassified as tracking, and their safety values propagate.}
  \label{fig:astrack}
  \vspace{-0.3cm}
\end{figure*}

\subsection{Web tracking detection}

ASTrack detects web tracking code by independently classifying the detected ASTs for each inspected URL. To this end, we initially label the URLs as tracking or not based on the most up-to-date filter lists. The classification is done by looking at the overall number of URLs that contain the identified functionality and are already known to perform tracking. The underlying idea is that, if one AST structure is mostly present in tracking URLs, its functionality is most probably used for tracking purposes. Thus, an unknown URL that shares the same functionality (i.e., the same AST) with other tracking URLs will also be a tracking URL, and we can automatically classify it as such. 

Fig.~\ref{fig:astrack} conceptually shows the process of the ASTrack automatic identification system during a usual browsing session. In the first step, the user opens a domain that loads a safe URL (i.e., not labeled as tracking). ASTrack automatically computes its inner AST identifiers. The circle represents the URL, and its inner hexagons are the ASTs internally loaded. As ASTrack does not have information about this URL, it will consider its inner ASTs as safe (green color and positive value). In the second step, a new safe URL is detected, and shares one AST with the previous URL. Their shared AST safety level becomes the sum of their individual safety levels, increasing it to 2. The third step introduces an already-known tracking URL. Its ASTs are considered unsafe (negative safety value and light red color) and consequently blocked. In the fourth step, a new tracking URL is detected, but this time it shares one AST with an URL that until now was considered safe. This shared AST safety value becomes zero. ASTrack does not have enough evidence to decide if it should be blocked or not. In this case, ASTrack allows its execution to maintain website functionality as much as possible. In the fifth step, a new tracking URL is detected and shares the same AST as the last URL. The safety value for the AST becomes negative. From now on, whenever the first URL is opened, ASTrack will block this specific AST but not the rest of them. In the sixth and seventh steps, new tracking URLs are detected sharing the same AST, thus decreasing its safety value again. The resultant safety value is considered enough evidence to autoclassify the AST as a tracking AST (red color). Consequently, each URL containing this piece of code is also a tracking URL. In the last step, ASTrack classifies the first safe URL as tracking. From now on, whenever it is detected, it will decrease the safety of all its internal ASTs propagating the information.

Performance-wise, ASTrack should be implementable inside the browser's internal JavaScript engine. The AST identifiers are simple enough to be computed during the initial DOM construction time. Once computed, comparing them to the subset of tracking ASTs is a \textit{set membership} type of problem, whose implementation can be done, for instance, by means of a \textit{bloom filter}. However, for practical reasons, in this research we will compute them and perform the comparison offline.

\BgThispage
\subsection{Web tracking removal}
Maintaining the functionality of the website is a key element for the adoption of any privacy protection method, as many users like the idea of improving their privacy but are discouraged by the associated website breakage. The high granularity obtained by ASTrack, which looks at the inner ASTs instead of the file itself, allows the method to minimize the functionality loss related to privacy protection purposes. In the worst-case scenario, for URLs with no shared ASTs, ASTrack performs equal to the filter lists. It blocks the complete code pertaining to that URL. However, if shared ASTs are present within the code, common as we will see in the next section, ASTrack automatically adapts to distinguish between them, selectively classifying tracking AST branches and URLs. For instance, in step 5 of Fig.~\ref{fig:astrack}, ASTrack has one safe URL within whose ASTs one is identified as unsafe. Thus, ASTrack can selectively prune the branch of this AST without compromising the rest of the code. This progressive detection achieves more detail than the traditional alternatives and minimizes the functionality loss usually suffered as a trade-off for privacy protection.

\BgThispage
\section{Evaluation}
\label{evaluation}
To evaluate ASTrack, we collected a labeled snapshot of all the URLs and resources pertaining to the top 10k most popular websites according to the Tranco List~\cite{tranco_list}. We used Selenium~\cite{jason_huggins_seleniumhq_2020} in combination with Mozilla Firefox and a customized version of uBlock Origin~\cite{hill_ublock_2020} to collect it.
Our customized uBlock intercepts all the HTTP requests, labels them according to its included lists, but allows them to pass through. To compute the ASTs, we used the JavaScript code parser Esprima~\cite{esprima}. For HTML files containing JavaScript, we automatically extract the code and compute its ASTs. Table~\ref{tab:tracking_dataset} contains information about the obtained data set. From the initial population of 10k websites, 8,179 domains were successfully inspected using a timeout of 60 seconds. The rest were not accessible at the time of the collection or made the timeout expire. The labeling process classified 41,274 JavaScript URLs as tracking. From the collected data, we precomputed the inner ASTs of each URL. More than 38\% of them are shared between different websites, validating our observation that many domains share common services.

\begin{table}
  \caption{Data set \& Evaluation}
  \label{tab:tracking_dataset}
  \resizebox{0.485\textwidth}{!}{
  \small
\begin{tabular}{lcc}
\hline
                                                           & \multicolumn{1}{l}{\textbf{Static evaluation}} & \multicolumn{1}{l}{\textbf{Dynamic evaluation}} \\ \hline
\textbf{Domains}                                           & \multicolumn{2}{c}{8,179}     \\
\textbf{URLs}                                              & \multicolumn{2}{c}{615,780}   \\
\textbf{JavaScript URLs}                                   & \multicolumn{2}{c}{161,593}   \\
\textbf{Tracking URLs}                                     & \multicolumn{2}{c}{41,274}    \\
\textbf{Unique ASTs}                                       & \multicolumn{2}{c}{7,015,542} \\
\textbf{Shared ASTs}                                       & \multicolumn{2}{c}{2,683,586} \\ \hline
\textbf{New tracking URLs}                                 & 3,409         & 2,183         \\
\textbf{New tracking JavaScript files}                     & 3,109         & 2,093         \\
\textbf{Tracking ASTs}                                     & 49,453        & 41,114        \\ \hline
\textbf{Precision}                                         & 98.52\%       & 98.47\%       \\ \hline
\end{tabular}}
\end{table}

With the collected data set, we feed ASTrack in order to classify the labeled ASTs and detect unknown tracking URLs. The evidence threshold, used to automatically classify an AST as tracking, was manually validated to maximize precision and minimize false positives that would break the functionality. Its value is dynamically computed as the 90\% of the total URLs containing the AST. Similarly, to avoid classifying ASTs without enough evidence, we empirically fixed a minimum of 10 different URLs containing the AST to propagate the tracking information. In order to evaluate ASTrack, we compare two different scenarios: a static evaluation with the complete interconnected graph before propagating the safety values, and a dynamic evaluation with no previous information about the ASTs available, forcing ASTrack to fill the graph.

\begin{figure}
  \centering
  \includegraphics[width=0.489\textwidth]{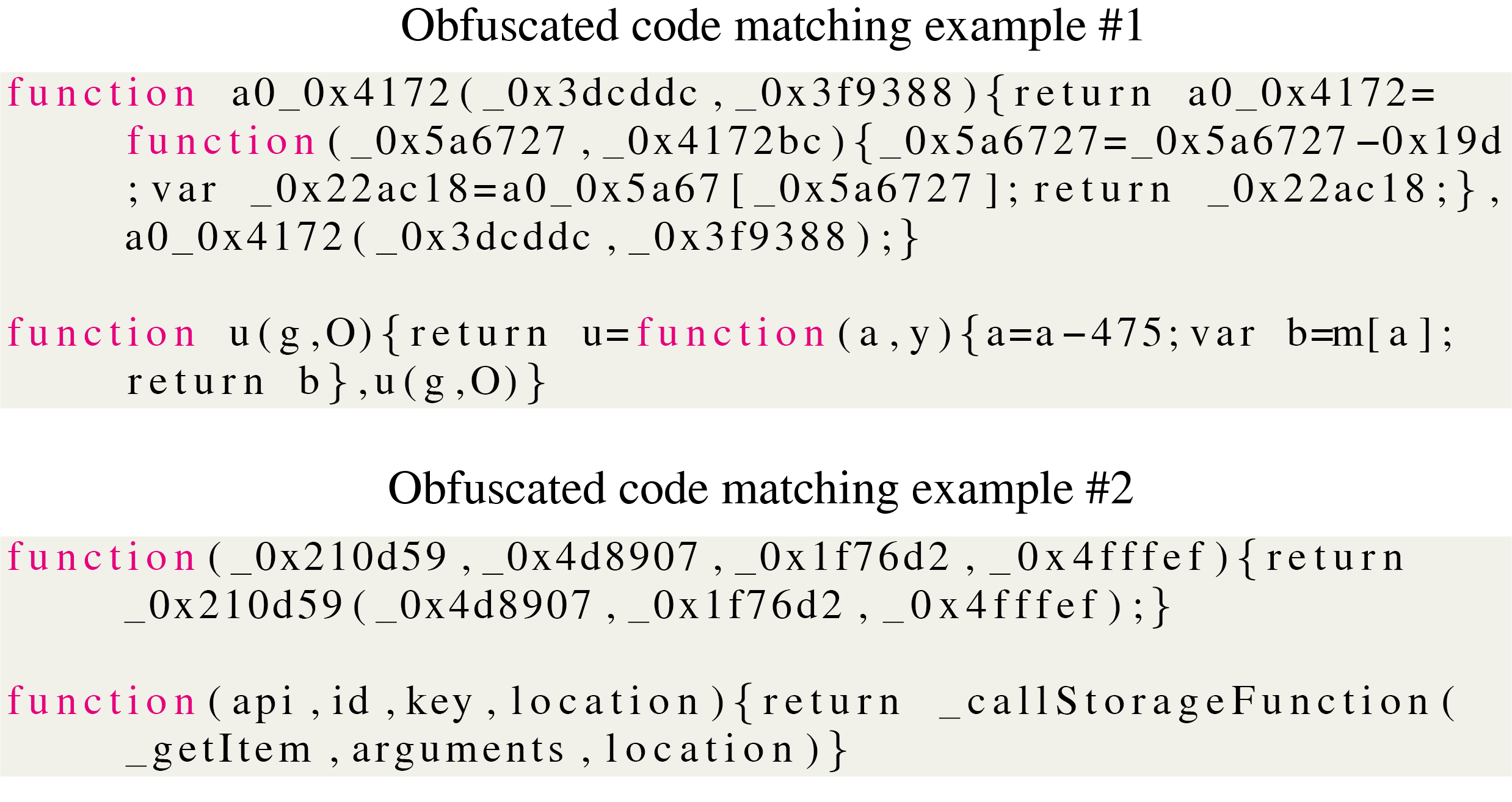}
  \caption{\textbf{Obfuscated code detection:} ASTrack was able to match obfuscated functions (top) with their alternative clear code (bottom) in different files by looking at their structure.}
  \label{fig:deobfuscation}
\end{figure}

\subsection{Static evaluation}
\label{best_case}
This experiment is composed of two phases. In the first phase, we create the complete graph, including all the relations and initial safety values for the shared ASTs. In the second phase, we run the ASTrack algorithm in order to identify tracking ASTs and propagate them to find new tracking URLs. This experiment is used to validate ASTrack's ability to find false negatives inside the filter lists. URLs not present in the filter lists but sharing a tracking AST (more than 90\% of the URLs containing it are known to be tracking) are most probably false negatives. Table~\ref{tab:tracking_dataset} includes the obtained results. ASTrack classified as tracking 49,453 ASTs and found 3,409 new tracking URLs, a 7.62\% increase with respect to the initially labeled data set. Inspecting those URLs, most of them pertained to the usual suspects (e.g., Google, Facebook, Twitter), while some others were files hosted in CDNs, which are hard to block without breaking page functionality on many websites. To validate them, we looked at the subset of files loaded by those URLs, composed of 3,109 JavaScript files. None of the files were already known to perform tracking (i.e., loaded by a different tracking URL). To study them, we first automatically checked their content for the inclusion of some frequent keywords used for tracking. The list of keywords is formed by the keywords previously found by Lerner et al. in~\cite{lerner_internet_2016} and by Iqbal et al. in~\cite{iqbal_fingerprinting_2021} for common stateless web tracking methods. We added some additional keywords for stateful tracking mechanisms (e.g. \textit{getCookie}, \textit{setCookie}, \textit{localStorage} or \textit{sessionStorage}). From the initial 3,109 files, 2,916 contained some of those keywords, with a median of 4 keywords per file. From the remaining 193 files, 50 were randomly selected and manually inspected. Approximately 76\% of them (38 out of 50) were recognized as tracking, while for the rest there was not enough evidence. Overall, the method presented more than 98\% of precision and automatically found about 3,400 false negatives. 

Three of the manually inspected files presented obfuscation techniques. ASTrack was able to automatically find structure coincidences with already known tracking files and correctly classify them. Fig.~\ref{fig:deobfuscation} shows some examples of obfuscated code and their alternative clear code found by ASTrack in different files. On the other hand, 1,564 (50.32\%) of the newly detected tracking files are webpackaged files. Webpack~\cite{noauthor_webpack_nodate} is a framework to easily pack different scripts inside only one file. It automatically looks for the needed dependencies and inserts them inside the file to allow for self-contained dynamic content. However, if the resource includes tracking libraries, privacy-protection tools cannot block it without suffering the functionality loss associated with blocking the rest of the included code. In contrast, ASTrack is able to detect inside them code portions used exclusively for tracking purposes.

\begin{lstlisting}[float=b, label=webpack_code, language=Caml, caption={Webpack files structure}, basicstyle=\scriptsize, numbers=none]
(window.webpackJsonp=window.webpackJsonp||[]).push([[261],{XygZ: ...
(self.webpackChunk_N_E=self.webpackChunk_N_E||[]).push([[4367],{4367: ...
(self.__LOADABLE_LOADED_CHUNKS__=self.__LOADABLE_LOADED_CHUNKS__||[]).push([[76429],{122954: ...
(window.webpackJsonpwebpackLogReporter=window.webpackJsonpwebpackLogReporter||[]).push([[5],{93: ...
\end{lstlisting}

\begin{figure}
  \centering
  \includegraphics[width=0.419\textwidth]{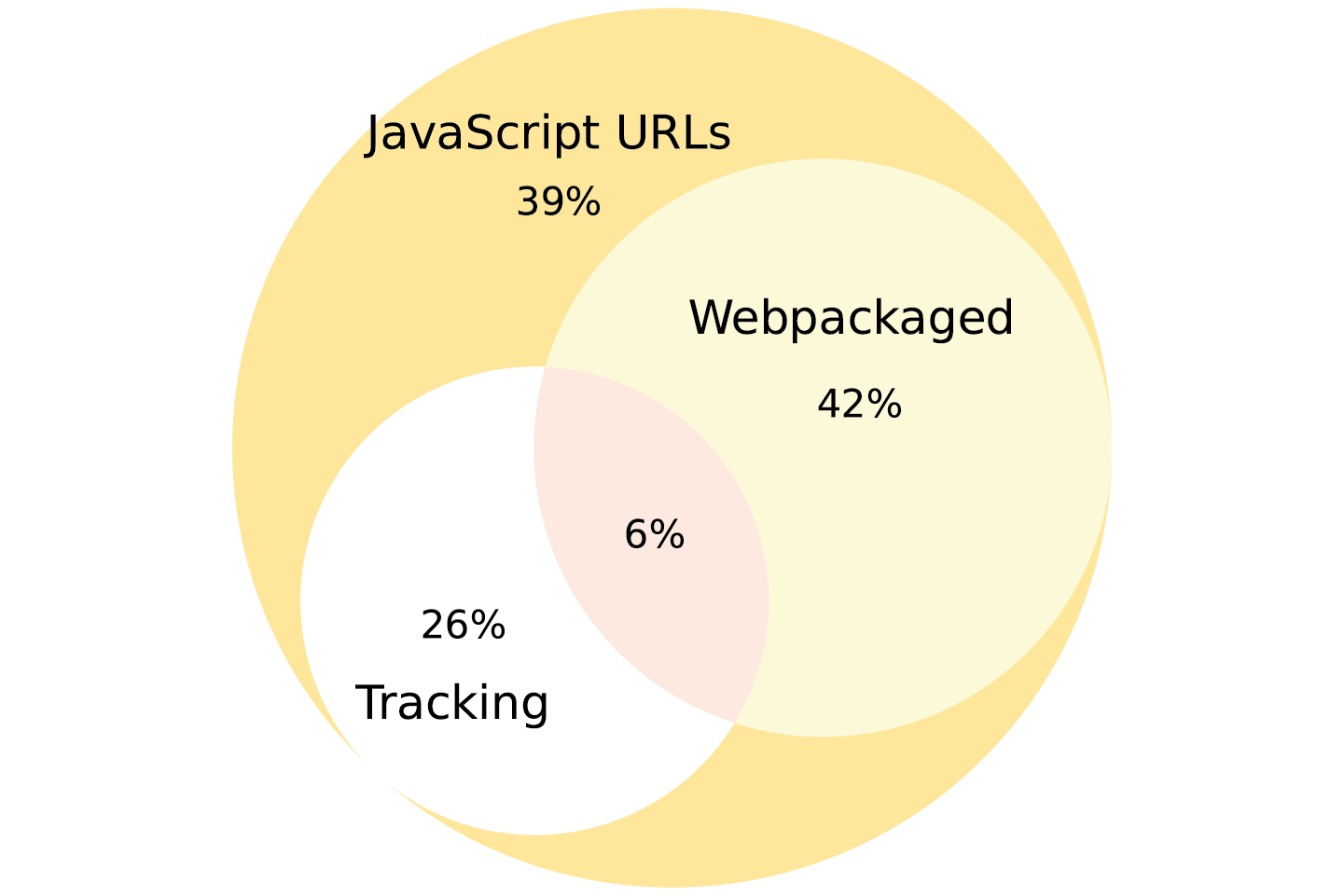}
  \caption{\textbf{Webpack and tracking presence:} 6\% of the JavaScript URLs contain tracking code inside webpackaged files. Traditional privacy-protecting tools suffer from functionality loss when blocked. ASTrack granularity allows us to selectively block only the tracking portions.}
  \label{fig:webpack_wild}
  \vspace{-0.3cm}
\end{figure}

\begin{figure}
  \centering
  \includegraphics[width=0.447\textwidth]{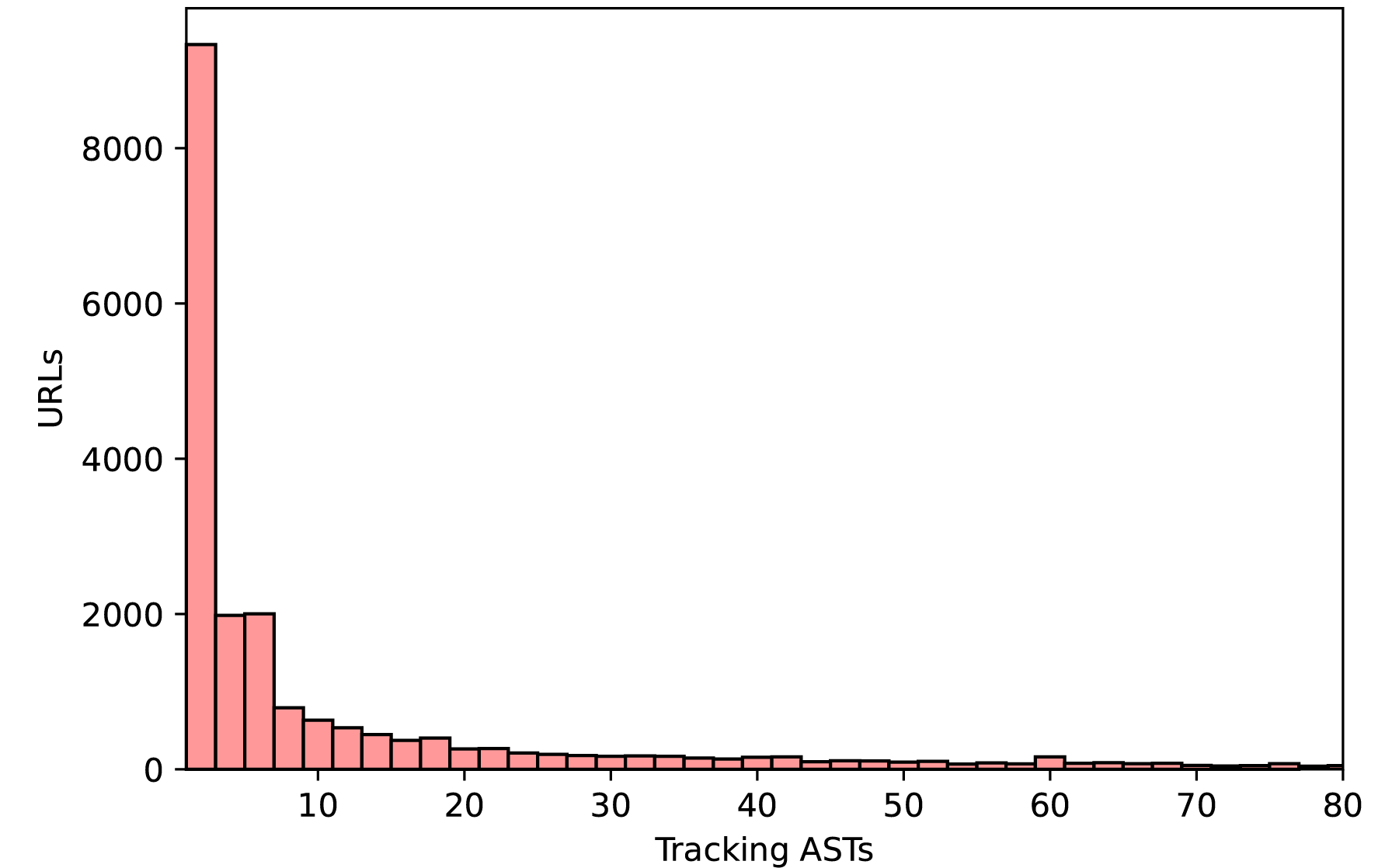}
  \caption{\textbf{Tracking AST distribution:} Distribution of the number of tracking ASTs included in tracking URLs. Most URLs contain only a few of them. Between all the ASTs included in the file (38 by median), ASTrack was able to discover the few ASTs whose code is used for tracking purposes. }
  \label{fig:ast_distribution}
  \vspace{-0.3cm}
\end{figure}

Listing~\ref{webpack_code} shows four examples of the first lines of webpackaged files. Although their code seems different, the structure of the code is always the same. Thus, thanks to ASTrack's ability to find shared AST structures, we can easily discover webpackaged files in the wild. To this end, we compared the initial portions of the AST identifiers of a subset of webpackaged files, and automatically extracted some patterns to identify them. Searching for those patterns in our collected AST data set, we accounted for the total number of webpackaged files and how many of them are classified as tracking. Fig.~\ref{fig:webpack_wild} shows the obtained results. About 39\% of files were safe, non-packaged files. On the other hand, approximately 48\% of the JavaScript files are webpackaged files, and 32\% include tracking. About a 6\% of the files are both webpackaged and include tracking. This represents 20\% of all the tracking URLs. Unfortunately, filter lists and other methods blocking complete resources will cause websites using those URLs to lose functionality. In contrast, ASTrack permits us to selectively block only the tracking ASTs while maintaining functionality in most cases. To discover if ASTrack is blocking complete files or only portions of them, we accounted for the number of tracking ASTs present in each tracking file. Fig.~\ref{fig:ast_distribution} shows the distribution of the number of tracking ASTs included inside tracking URLs. According to our results, more than 60\% of the detected tracking files include only one or two tracking ASTs (the median is 6). In comparison, the median of ASTs per file, including non-tracking ones, is 38 ASTs per file (35 for non-tracking URLs). Thus, ASTrack is selectively classifying only the branches identifying web tracking. 

\BgThispage
\subsection{Dynamic evaluation}
\label{worst_case}
Our second experiment allows us to evaluate the adaptation properties of ASTrack's algorithm. We run the system with the AST graph empty. Only the URL tracking information in the filter lists is available. In this case, we can study if, by applying the method to a common browsing session, the system is able to gradually discover new web tracking and correctly identify tracking ASTs. To this end, we consecutively feed ASTrack with data from one URL at a time. This forces the algorithm to progressively compute the connections between the ASTs loaded and their safety values. The URL insertion order is set to match the website rank of the Tranco List. As it is based on domain popularity, it is a good representation of a common browsing session, where pages that are very popular are more likely to be accessed before websites with a lower rank. For each domain, its accessed URLs are introduced arbitrarily, as online resources are mostly loaded asynchronously.

Table~\ref{tab:tracking_dataset} includes the obtained results. After the 8,179 domain insertions, ASTrack labeled as tracking 41,114 ASTs. This represents a reduction of about 16\% in comparison to the static evaluation process presented in the last section. However, ASTrack found 43,457 tracking URLs, a decrease of less than 3\% with respect to the complete model approach. Between them, 2,183 (5\%) were not previously included in the initial filter lists. Once more, we studied the 2,093 JavaScript files loaded by those URLs. Interestingly, comparing them with the subset of files found during the static evaluation, about 35\% of them were new resources, not previously classified as tracking. This highlights the main weakness of the filter lists: good precision but low recall. Many URLs incorrectly classified as safe can increase the overall safety value of their inner ASTs and not be detected as tracking using the complete graph. In contrast, progressively feeding ASTrack can help find them. Following the same methodology, we automatically checked for the inclusion of frequent tracking keywords inside the files. 1,868 files included at least one of them, with a median of 5 keywords per file. From the subset of files that do not include keywords and were not detected during the complete graph experiment, we randomly selected 50 files and manually inspected them. The inspection discovered only seven of them mistakenly classified as tracking (mainly \textit{reddit.com} static scripts). Finally, in line with our previous finding, 46.67\% of them were webpackaged files. Overall, as in the static evaluation, more than 98\% of the detected URLs were correctly classified, validating the adaptation properties of ASTrack to automatically discover web tracking.

\BgThispage
\subsection{Tracking removal and website breakage}
\label{tracking_removal}

To evaluate the tracking removal efficiency of ASTrack we measure the functionality loss associated with it. In this work, for practical reasons, we use file replacements to test and validate our new web tracking removal methodology, similar to the proposal in~\cite{smith_sugarcoat_2021}. To generate the file replacements, we automatically remove all the code pertaining to tracking ASTs from the tracking files detected during the static and dynamic evaluations (25,840 files). Unfortunately, although web technologies have been around for about 30 years, there is not yet a defined method to evaluate website breakage. Proposals such as \cite{choudhary_detecting_2011} and \cite{mesbah_automated_2011} were focused on breakage due to JavaScript engine rendering differences, looking at DOM discrepancies. However, cleaning web tracking systems modifies the DOM structure of the website, but may not deteriorate its functionality as they are additional systems that are usually not related to the website content. Until now, subjective manual analysis has been used to detect functionality loss (e.g. \cite{iqbal_adgraph_nodate, iqbal_fingerprinting_2021, smith_sugarcoat_2021}). However, this approach does not scale, limiting the number of evaluated websites to the manual labor you can afford (a few dozens in previous works). In this work, we introduce a new alternative methodology, using computer vision techniques in order to discover website breakage suspicious websites prior to the manual inspection.

The idea is to compute the similarity between screenshots taken with and without the modifications introduced in our process. Note that, as many websites include animations and other dynamically modified content, we need to take care of the expected variability of a website. If the obtained screenshots are not similar enough, we consider them suspicious of functionality loss. The proposed process is:

\begin{enumerate}
    \item Collect multiple \textbf{independent screenshot data sets from two vanilla browsers in pairs} for the desired population. Each pair has to be collected in parallel to minimize the impact of external events between them (e.g., network congestion, periodic maintenance).
    \item Collect \textbf{one more data set by replacing one vanilla browser with our modified approach}. In our case, we will use the file replacements created by removing the branches found in the last section.
    \item Compare the obtained pairs of screenshot data sets using \textbf{Normalized Cross Correlation} (NCC)~\cite{zhao_image_2006} or another equivalent technique to detect the percentage of similarity between them.
    \item Compute the \textbf{expected similarity deviation per website} by means of the standard deviation and its confidence intervals between the multiple similarity measures obtained from the vanilla browsers.
    \item Check if our \textbf{modified output similarity is within the expected deviation} for each website.
    \item For websites that are not similar enough, \textbf{generate a diff file between both screenshots}, highlighting the pixels that are different between them. In a posterior phase, an expert can visually inspect the diff file along with the screenshots to better classify the suspicious websites.
\end{enumerate}

\begin{figure}
  \centering
  \includegraphics[width=0.489\textwidth]{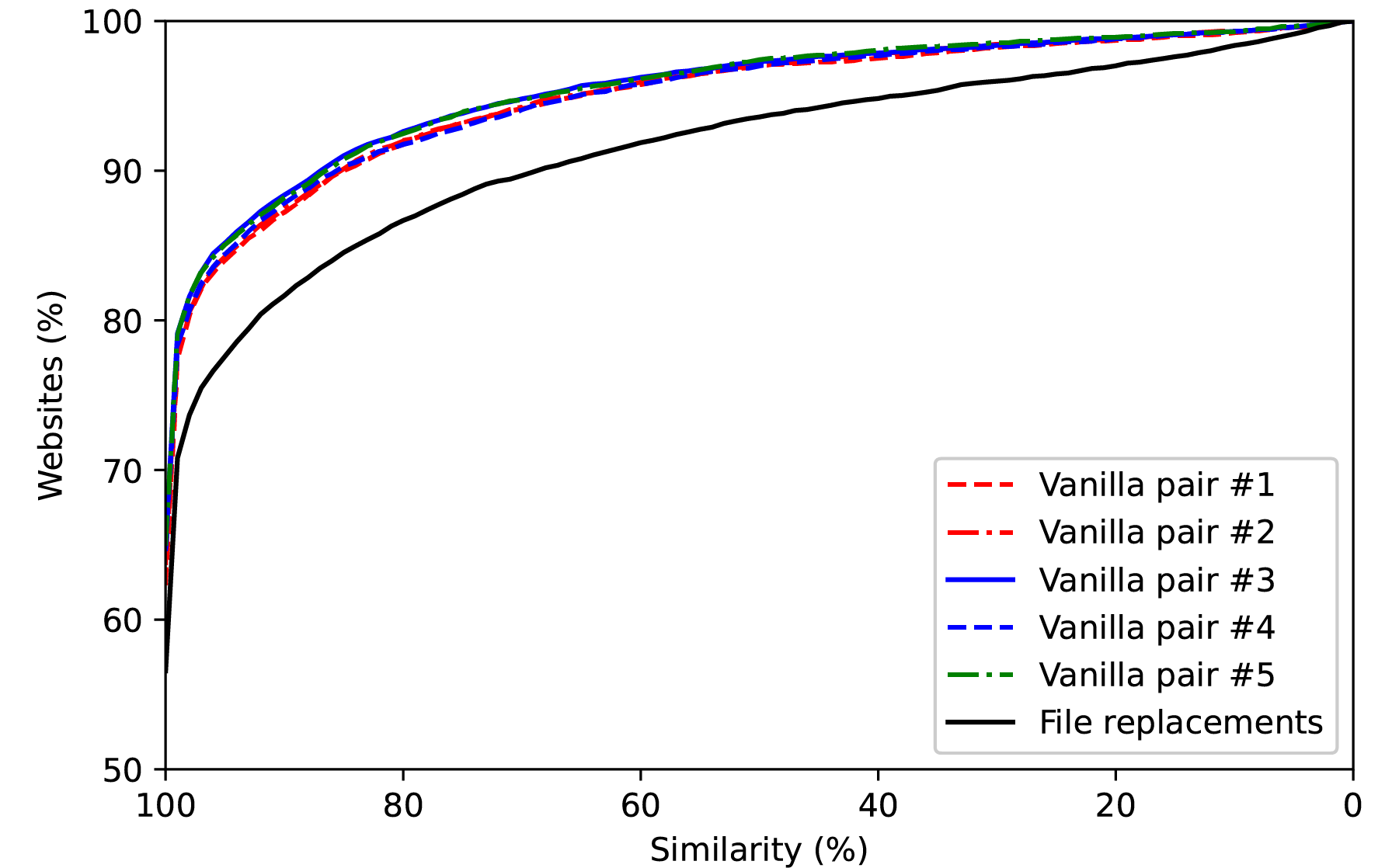}
  \caption{\textbf{Similarity comparison (CDF):} Distribution of websites in function of their visual similarity percentage (Normalized Cross Correlation). Five data sets comparing two vanilla browsers and one more comparing a vanilla browser with our file replacements were taken.} 
  \label{fig:similarity}
  \vspace{-0.3cm}
\end{figure}

Applying this methodology, we collected a new data set of the top 10k most popular websites with one vanilla browser alongside another browser using our file replacements. The resulting crawl contains information about 8,050 Internet domains, and our plugin replaced almost 23k elements with their clean versions. Surprisingly, although we are not actively blocking URLs, the vanilla browser loaded about 22k more than our system. Thus, this excess of traffic can be directly attributed to tracking purposes. Approximately 72\% of domains (5,751 out of 8,050) benefited from a privacy improvement thanks to our removal system. On average, 4.05 JavaScript files were replaced in websites where tracking was detected. This represents a median of 62\% tracking reduction in comparison with the vanilla browser.

Next, we collected five independent screenshot data sets of two vanilla browsers in parallel for the same 10k websites. We used ImageMagick~\cite{imagemagick_compare} to compute the NCC similarity values and the diff files between all the pairwise obtained data sets. Finally, we computed the expected similarity deviation as well as the 95\% confidence intervals for each website from the measures collected in the vanilla comparisons. Fig.~\ref{fig:similarity} shows the cumulative distribution function (CDF) values for each of the pairwise data set comparisons. The difference obtained by comparing two vanilla browsers is very similar for all five collected data sets. Only about 62\% of websites are completely equal between both vanilla browsers, with another 30\% of them having similarity values higher than 80\%. The remaining 8\% present a difference bigger than 20\% due to dynamic content and animations. In contrast, in our privacy-friendly browser, there are about 6\% to 7\% of websites that present similarity values lower than their counterparts in the vanilla browser. Overall, looking at the expected deviation, 1,753 websites using file replacements (21.7\%) were classified as suspicious of website breakage. 

\begin{figure}
  \centering
  \includegraphics[width=0.449\textwidth]{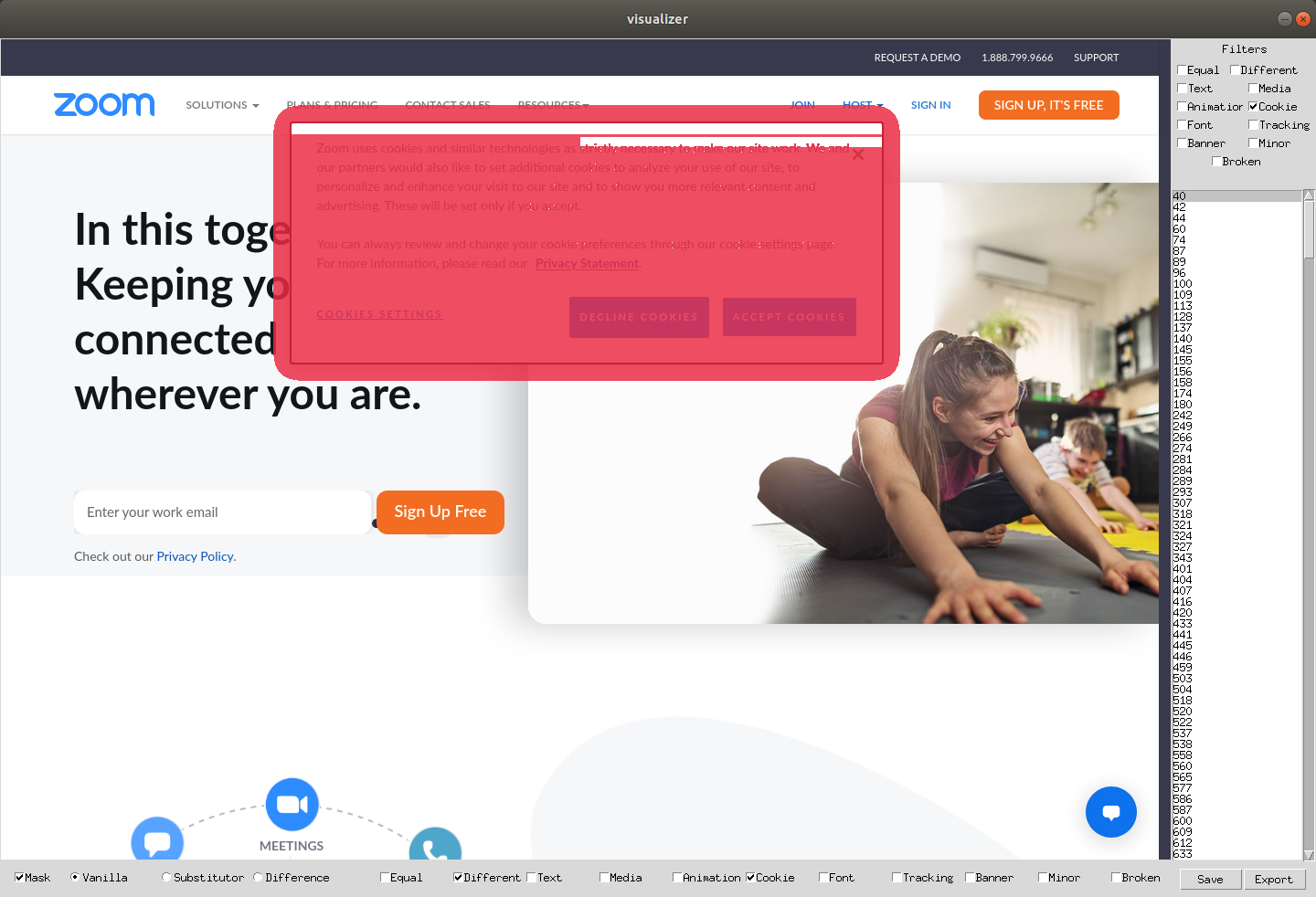}
  \caption{\textbf{Visualizer:} Self-developed visualization tool to easily inspect and compare screenshots. A diff mask can be applied to highlight the differences (red pixels).}
  \label{fig:visualizer}
  \vspace{-0.3cm}
\end{figure}

In order to explore them, we developed a small visualization tool that allows us to easily switch between the screenshots. It also allows us to imprint the pixel differences between them as a mask over any of the pictures. Fig.~\ref{fig:visualizer} shows a screenshot of the tool. We included a set of checkboxes to easily classify the reason for the similarity gap and also filters to group them by type. Using the tool, we were able to inspect all the suspicious websites in less than one day. The classification is divided into 9 groups, highlighting the main reasons for the screenshot difference:

\BgThispage
\begin{itemize}
    \item \textbf{Animation}: Elements with animations or changing at defined intervals, such as timed sliding banners.
    \item \textbf{Banner}: Advertisements or other banners.
    \item \textbf{Broken}: Functionality loss or website breakage. 
    \item \textbf{Cookie}: Missing cookie banners.
    \item \textbf{Fonts}: The original font is not available, and a default one is used instead (minimal impact).
    \item \textbf{Media}: Dynamically modified media content (e.g., videos, pictures, logos, icons).
    \item \textbf{Minor}: Minor dynamic content that usually varies with time, such as clocks, numbers of visits, views, etc.
    \item \textbf{Text}: Dynamically modified text content.
    \item \textbf{Tracking}: Visually visible tracking elements such as missing social network icons, captchas, anti-adblockers, or country detection pop-ups.
\end{itemize}

The only category considered as website breakage is the ``broken'' one. Note that it does not only contain functionality loss but also usability problems and aesthetically unpleasant modifications that would be obvious for the common user (e.g. missing icons, pictures, or broken animations). Websites have been classified in more than one group when needed. In particular, some anti-adblocking systems detected our system and blocked the website. In this case, the website has been considered both, ``tracking'' and ``broken''.  

\begin{figure}
  \centering
  \includegraphics[width=0.449\textwidth]{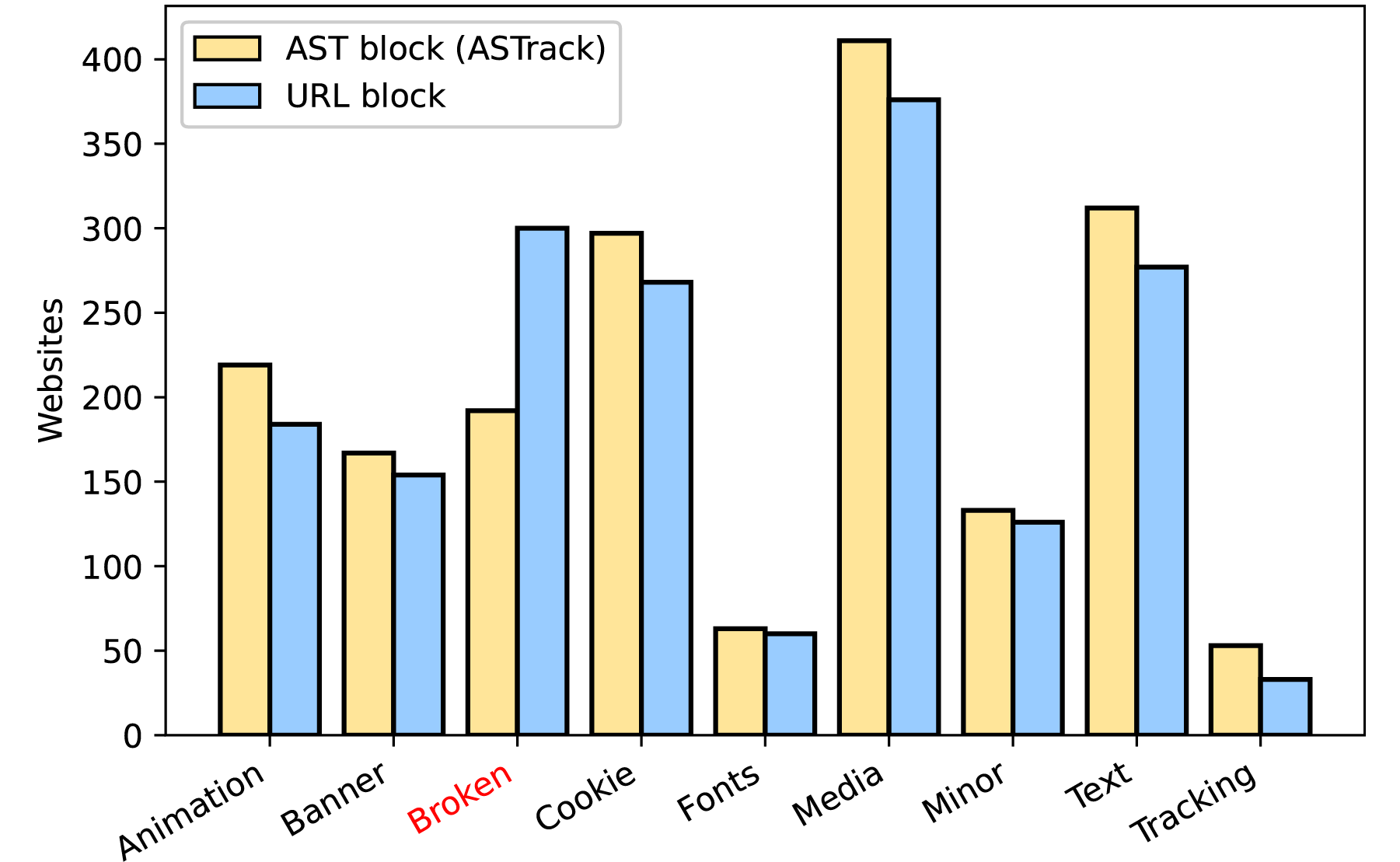}
  \caption{\textbf{Visual difference reason:} Main reason for the similarity gap between the vanilla browser and ASTrack (yellow). The figure also includes the same results but blocking URLs instead of only ASTs (blue). ASTrack reduces the broken websites by a 36\% (192 vs. 301) removing only the tracking AST and maintaining the rest of the file intact.}
  \label{fig:classification}
  \vspace{-0.3cm}
\end{figure}

\begin{figure*}
  \centering
  \includegraphics[width=0.94\textwidth]{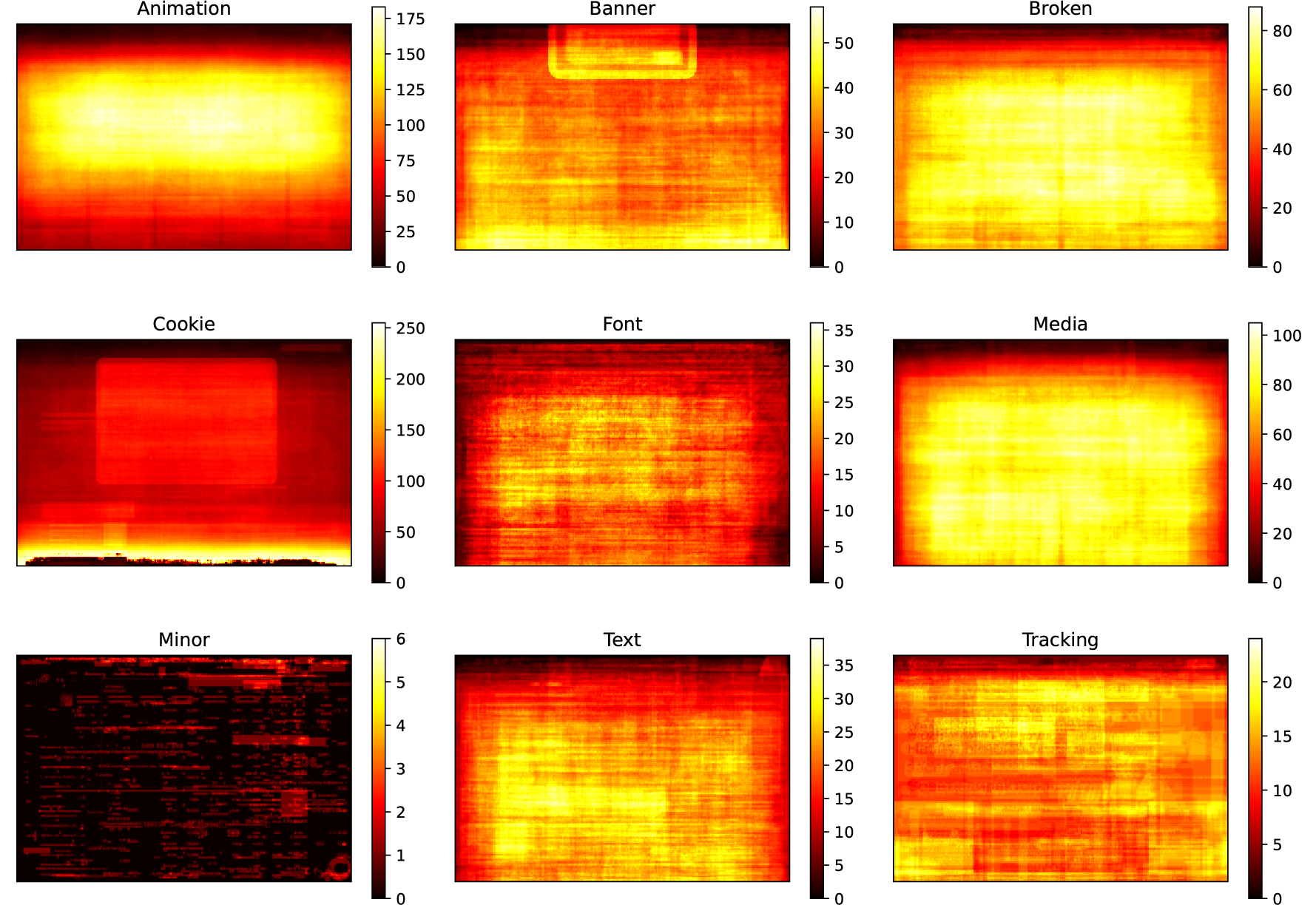}
  \caption{\textbf{Difference heatmaps:} Heatmaps containing the pixel distribution of the difference between the vanilla browser and our system. Some of them, such as the \textit{cookie} or \textit{banner} removal heatmap, present clear patterns. \textit{Animation} heatmap highlights the usual main sliding-banner position. Others, like \textit{media} an \textit{broken} heatmaps, share many characteristics.}
  \label{fig:heatmaps}
  \vspace{-0.3cm}
\end{figure*}

Fig.~\ref{fig:classification} shows the classification of the 1,753 suspicious websites. In order to compare the results with the traditional tools, the figure also includes the classification for the same subset of websites, but blocking URLs instead of using the files cleaned by ASTrack. The results show that using ASTs to selectively remove tracking code decreases the number of broken websites by 36\% (192 pages vs 301 pages). Most of the websites broken by blocking URLs but functional by blocking ASTs include broken animations (43.7\%) or missing media files (39.4\%). Overall, using ASTrack only 10.9\% of the suspicious websites presented functionality loss or visual problems. The rest were mostly due to dynamic content modifications (60.58\%), missing cookie banners (16.77\%) or different fonts (3.59\%). Fig.~\ref{fig:heatmaps} contains heatmaps highlighting the visual difference distribution for each of the categories. The figure presents clear patterns for the ``cookie'' and ``banner'' positioning as well as for the animation category, which is mainly composed of sliding animations in the center of the page. As expected, dynamic ``text'' and ``media'' are mostly distributed in the content portion of the website. ``Minor'' category only has a few instances, making them very recognizable. In total, ASTrack functionality loss was only present in 2.3\% of websites (192 out of 8,050). 

\section{Conclusions and Future Work}
\label{conclusions}
In this work, we presented ASTrack, a new methodology to detect and selectively remove web tracking systems. To this end, it uses an abstraction of the website's JavaScript AST structure that represents its functionality. The method works by identifying tracking functionality shared by multiple websites. The abstraction from the actual code makes the system robust against obfuscation and other similar renaming techniques, a common problem in many other solutions. Moreover, the high granularity achieved by the method allows the system to automatically prune only the code pieces exclusively dedicated to tracking purposes, minimizing the functionality loss. We also presented a new methodology to discover website breakage that compares the visual differences of the websites in order to highlight those suspicious of being broken.

Our results show that ASTrack presents a detection precision higher than 98\%. Moreover, thanks to its adaptability, in the evaluation of the top 10k most popular websites, ASTrack found more than 3,400 new tracking URLs (7.62\% increase) and identified almost 50k tracking ASTs. Using the selective tracking removal to clean tracking files, ASTrack achieved a 62\% tracking removal in more than 72\% of the websites. Moreover, it also obtained a 36\% decrease in functionality loss in comparison with the filter lists. We estimate that almost 98\% of the websites maintained full functionality due to the high granularity of working with AST's structure, the main contribution of this work. Our future work includes improving the ASTrack tracking removal system to intelligently substitute branches instead of completely removing them, decreasing even more the functionality loss. We also plan to improve web tracking detection by studying simple AST transformations useful to find mostly equivalent ASTs but presenting small differences in their structure. Finally, we also expect to improve the website breakage detection method by applying ML to automatically classify the visual differences of suspicious websites using the obtained heatmaps and data sets. 
The data sets and source code of ASTrack are publicly available at~\cite{ismael_castell-uroz_online_2020}.

\BgThispage
\section{Acknowledgments}
This publication is part of the Spanish I+D+i project TRAINER-A (ref.~PID2020-118011GB-C21), funded by MCIN/ AEI/10.13039/501100011033. This work is also partially supported by the NII internship program.

\bibliographystyle{ieeetr}
\balance

\bibliography{references}

\begin{thebibliography}{10}

\bibitem{acar_web_2014}
G.~Acar, C.~Eubank, S.~Englehardt, M.~Juarez, A.~Narayanan, and C.~Diaz, ``The
  {Web} {Never} {Forgets}: {Persistent} {Tracking} {Mechanisms} in the
  {Wild},'' in {\em Proceedings of the 2014 {ACM} {SIGSAC} {Conference} on
  {Computer} and {Communications} {Security}}, {CCS} '14, (Scottsdale, Arizona,
  USA), pp.~674--689, Association for Computing Machinery, Nov. 2014.

\bibitem{englehardt_cookies_2015}
S.~Englehardt, D.~Reisman, C.~Eubank, P.~Zimmerman, J.~Mayer, A.~Narayanan, and
  E.~W. Felten, ``Cookies {That} {Give} {You} {Away}: {The} {Surveillance}
  {Implications} of {Web} {Tracking},'' in {\em Proceedings of the 24th
  {International} {Conference} on {World} {Wide} {Web}}, {WWW} '15, (Florence,
  Italy), pp.~289--299, International World Wide Web Conferences Steering
  Committee, May 2015.

\bibitem{li_trackadvisor_2015}
T.-C. Li, H.~Hang, M.~Faloutsos, and P.~Efstathopoulos, ``{TrackAdvisor}:
  {Taking} {Back} {Browsing} {Privacy} from {Third}-{Party} {Trackers},'' in
  {\em Passive and {Active} {Measurement}} (J.~Mirkovic and Y.~Liu, eds.),
  Lecture {Notes} in {Computer} {Science}, (Cham), pp.~277--289, Springer
  International Publishing, 2015.

\bibitem{metwalley_unsupervised_2015}
H.~Metwalley, S.~Traverso, and M.~Mellia, ``Unsupervised {Detection} of {Web}
  {Trackers},'' in {\em 2015 {IEEE} {Global} {Communications} {Conference}
  ({GLOBECOM})}, pp.~1--6, Dec. 2015.

\bibitem{nikiforakis_cookieless_2013}
N.~Nikiforakis, A.~Kapravelos, W.~Joosen, C.~Kruegel, F.~Piessens, and
  G.~Vigna, ``Cookieless {Monster}: {Exploring} the {Ecosystem} of
  {Web}-{Based} {Device} {Fingerprinting},'' in {\em 2013 {IEEE} {Symposium} on
  {Security} and {Privacy}}, pp.~541--555, May 2013.

\bibitem{lerner_internet_2016}
A.~Lerner, A.~K. Simpson, T.~Kohno, and F.~Roesner, ``Internet {Jones} and the
  {Raiders} of the {Lost} {Trackers}: {An} {Archaeological} {Study} of {Web}
  {Tracking} from 1996 to 2016,'' in {\em 25th \{USENIX\} Security Symposium
  (\{USENIX\} Security 16)}, USENIX Association, 2016.

\bibitem{iqbal_fingerprinting_2021}
U.~Iqbal, S.~Englehardt, and Z.~Shafiq, ``Fingerprinting the {Fingerprinters}:
  {Learning} to {Detect} {Browser} {Fingerprinting} {Behaviors},'' in {\em 2021
  {IEEE} {Symposium} on {Security} and {Privacy} ({SP})}, pp.~1143--1161, May
  2021.

\bibitem{castell-uroz_tracksign_2021}
I.~Castell-Uroz, J.~Solé-Pareta, and P.~Barlet-Ros, ``{TrackSign}: {Guided}
  {Web} {Tracking} {Discovery},'' in {\em {IEEE} {INFOCOM} 2021 - {IEEE}
  {Conference} on {Computer} {Communications}}, pp.~1--10, May 2021.

\bibitem{hill_ublock_2020}
R.~Hill, ``{uBlock} {Origin},'' July 2022.
\newblock https://github.com/gorhill/uBlock.

\bibitem{adblock_plus_adblock_2020}
{AdBlock Plus}, ``Adblock {Plus},'' July 2022.
\newblock https://adblockplus.org/en/.

\bibitem{wu_machine_2016}
Q.~Wu, Q.~Liu, Y.~Zhang, P.~Liu, and G.~Wen, ``A {Machine} {Learning}
  {Approach} for {Detecting} {Third}-{Party} {Trackers} on the {Web},'' in {\em
  Computer {Security} – {ESORICS} 2016} (I.~Askoxylakis, S.~Ioannidis,
  S.~Katsikas, and C.~Meadows, eds.), Lecture {Notes} in {Computer} {Science},
  (Cham), pp.~238--258, Springer International Publishing, 2016.

\bibitem{ikram_towards_2017}
M.~Ikram, H.~J. Asghar, M.~A. Kaafar, A.~Mahanti, and B.~Krishnamurthy,
  ``Towards {Seamless} {Tracking}-{Free} {Web}: {Improved} {Detection} of
  {Trackers} via {One}-class {Learning},'' {\em Proceedings on Privacy
  Enhancing Technologies}, vol.~2017, pp.~79--99, Jan. 2017.

\bibitem{krishnamurthy_measuring_2007}
B.~Krishnamurthy, D.~Malandrino, and C.~E. Wills, ``Measuring privacy loss and
  the impact of privacy protection in web browsing,'' in {\em Proceedings of
  the 3rd symposium on {Usable} privacy and security}, {SOUPS} '07,
  (Pittsburgh, Pennsylvania, USA), pp.~52--63, Association for Computing
  Machinery, July 2007.

\bibitem{mazel_comparison_2019}
J.~Mazel, R.~Garnier, and K.~Fukuda, ``A comparison of web privacy protection
  techniques,'' {\em Computer Communications}, vol.~144, pp.~162--174, Aug.
  2019.

\bibitem{iqbal_adgraph_nodate}
U.~Iqbal, P.~Snyder, S.~Zhu, B.~Livshits, Z.~Qian, and Z.~Shaﬁq, ``{ADGRAPH}:
  {A} {Graph}-{Based} {Approach} to {Ad} and {Tracker} {Blocking},'' {\em IEEE
  Symposium on Security and Privacy 2020}, p.~14, 2019.

\bibitem{smith_sugarcoat_2021}
M.~Smith, P.~Snyder, B.~Livshits, and D.~Stefan, ``{SugarCoat}:
  {Programmatically} {Generating} {Privacy}-{Preserving}, {Web}-{Compatible}
  {Resource} {Replacements} for {Content} {Blocking},'' in {\em Proceedings of
  the 2021 {ACM} {SIGSAC} {Conference} on {Computer} and {Communications}
  {Security}}, {CCS} '21, (New York, NY, USA), pp.~2844--2857, Association for
  Computing Machinery, Nov. 2021.

\bibitem{implementing_web_tracking}
G.~Fleischer, ``Implementing web tracking,'' {\em Proc. Black Hat USA Conf.
  Briefings}, pp.~1--37, July 2012.

\bibitem{bursztein_tracking_nodate}
E.~Bursztein, ``Tracking users that block cookies with a http redirect,'' July
  2011.
\newblock
  https://elie.net/blog/security/tracking-users-that-block-cookies-with-a-http-redirect/.

\bibitem{grossman_tracking_nodate}
J.~Grossman, ``Tracking users with {Basic} {Auth},'' Apr. 2007.
\newblock
  https://blog.jeremiahgrossman.com/2007/04/tracking-users-without-cookies.html.

\bibitem{ayenson_flash_2011}
M.~D. Ayenson, D.~J. Wambach, A.~Soltani, N.~Good, and C.~J. Hoofnagle, ``Flash
  {Cookies} and {Privacy} {II}: {Now} with {HTML5} and {ETag} {Respawning},''
  July 2011.
\newblock https://papers.ssrn.com/abstract=1898390.

\bibitem{gdpr}
``{General Data Protection Regulation},'' May 2022.
\newblock https://gdpr.eu/.

\bibitem{ccpa}
``{California Consumer Privacy Act},'' May 2022.
\newblock https://www.oag.ca.gov/privacy/ccpa.

\bibitem{standford_pipl_2021}
S.~University, ``{Translation}: {Personal} {Information} {Protection} {Law} of
  the {People’s} {Republic} of {China},'' Oct. 2021.
\newblock
  https://digichina.stanford.edu/work/translation-personal-information-protection-law-of-the-peoples-republic-of-china-effective-nov-1-2021/.

\bibitem{safari_builtin}
J.~Wilander, ``Full third-party cookie blocking and more,'' Mar. 2020.
\newblock
  https://webkit.org/blog/10218/full-third-party-cookie-blocking-and-more/.

\bibitem{firefox_builtin}
M.~Wood, ``Today's {Firefox} blocks third-party tracking cookies and
  cryptomining by default,'' 2019.
\newblock
  https://blog.mozilla.org/products/firefox/todays-firefox-blocks-third-party-tracking-cookies-and-cryptomining-by-default/.

\bibitem{google_cookies}
``An updated timeline for {Privacy} {Sandbox} milestones,'' June 2021.

\bibitem{boda_user_2012}
K.~Boda, {\'A}.~M. F{\"o}ldes, G.~G. Guly{\'a}s, and S.~Imre, ``User tracking
  on the web via cross-browser fingerprinting,'' in {\em Information Security
  Technology for Applications} (P.~Laud, ed.), (Berlin, Heidelberg),
  pp.~31--46, Springer Berlin Heidelberg, 2012.

\bibitem{unger_shpf_2013}
T.~Unger, M.~Mulazzani, D.~Frühwirt, M.~Huber, S.~Schrittwieser, and
  E.~Weippl, ``{SHPF}: {Enhancing} {HTTP}({S}) {Session} {Security} with
  {Browser} {Fingerprinting},'' in {\em 2013 {International} {Conference} on
  {Availability}, {Reliability} and {Security}}, pp.~255--261, Sept. 2013.

\bibitem{fifield_fingerprinting_2015}
D.~Fifield and S.~Egelman, ``Fingerprinting {Web} {Users} {Through} {Font}
  {Metrics},'' in {\em Financial {Cryptography} and {Data} {Security}}
  (R.~Böhme and T.~Okamoto, eds.), Lecture {Notes} in {Computer} {Science},
  (Berlin, Heidelberg), pp.~107--124, Springer, 2015.

\bibitem{snyder_browser_2016}
P.~Snyder, L.~Ansari, C.~Taylor, and C.~Kanich, ``Browser {Feature} {Usage} on
  the {Modern} {Web},'' in {\em Proceedings of the 2016 {Internet}
  {Measurement} {Conference}}, {IMC} '16, (Santa Monica, California, USA),
  pp.~97--110, Association for Computing Machinery, Nov. 2016.

\bibitem{mowery2012pixel}
K.~Mowery and H.~Shacham, ``Pixel perfect: Fingerprinting canvas in html5,''
  {\em Proceedings of W2SP}, vol.~2012, 2012.

\bibitem{englehardt_online_2016}
S.~Englehardt and A.~Narayanan, ``Online {Tracking}: {A} 1-million-site
  {Measurement} and {Analysis},'' in {\em Proceedings of the 2016 {ACM}
  {SIGSAC} {Conference} on {Computer} and {Communications} {Security}}, {CCS}
  '16, pp.~1388--1401, Association for Computing Machinery, Oct. 2016.

\bibitem{cao2017cross}
Y.~Cao, S.~Li, E.~Wijmans, {\em et~al.}, ``(cross-) browser fingerprinting via
  os and hardware level features.,'' in {\em Proceedings 2017 {Network} and
  {Distributed} {System} {Security} {Symposium}}, Internet Society, 2017.

\bibitem{starov_xhound_2017}
O.~Starov and N.~Nikiforakis, ``{XHOUND}: {Quantifying} the
  {Fingerprintability} of {Browser} {Extensions},'' in {\em 2017 {IEEE}
  {Symposium} on {Security} and {Privacy} ({SP})}, pp.~941--956, May 2017.

\bibitem{zhu_eluding_2021}
S.~Zhu, Z.~Wang, X.~Chen, S.~Li, K.~Man, U.~Iqbal, Z.~Qian, K.~S. Chan, S.~V.
  Krishnamurthy, Z.~Shafiq, Y.~Hao, G.~Li, Z.~Zhang, and X.~Zou, ``Eluding
  {ML}-based {Adblockers} {With} {Actionable} {Adversarial} {Examples},'' in
  {\em Annual {Computer} {Security} {Applications} {Conference}}, {ACSAC} '21,
  (New York, NY, USA), pp.~541--553, Association for Computing Machinery, Dec.
  2021.

\bibitem{noauthor_webpack_nodate}
``webpack,'' July 2022.
\newblock https://webpack.js.org/.

\bibitem{open_source_easylist_2020}
``{EasyList},'' July 2022.
\newblock https://easylist.to/easylist/easylist.txt.

\bibitem{open_source_easyprivacy_2020}
``{EasyPrivacy},'' July 2022.
\newblock https://easylist.to/easylist/easyprivacy.txt.

\bibitem{firefox_fingerprinting}
``Firefox's protection against fingerprinting,'' July 2022.
\newblock
  https://support.mozilla.org/en-US/kb/firefox-protection-against-fingerprinting.

\bibitem{tor_fingerprinting}
``Browser {Fingerprinting}: {An} {Introduction} and the {Challenges} {Ahead}
  {\textbar} {Tor} {Project},'' Sept. 2019.
\newblock
  https://blog.torproject.org/browser-fingerprinting-introduction-and-challenges-ahead/.

\bibitem{tranco_list}
``A research-oriented top sites ranking hardened against manipulation -
  {Tranco},'' May 2022.
\newblock https://tranco-list.eu/list/W9359/1000000.

\bibitem{jason_huggins_seleniumhq_2020}
{Jason Huggins}, ``{SeleniumHQ} {Browser} {Automation},'' Feb. 2020.

\bibitem{esprima}
``Esprima.''
\newblock https://esprima.org/.

\bibitem{choudhary_detecting_2011}
S.~R. Choudhary, ``Detecting cross-browser issues in web applications,'' in
  {\em 2011 33rd {International} {Conference} on {Software} {Engineering}
  ({ICSE})}, pp.~1146--1148, May 2011.

\bibitem{mesbah_automated_2011}
A.~Mesbah and M.~R. Prasad, ``Automated cross-browser compatibility testing,''
  in {\em Proceedings of the 33rd {International} {Conference} on {Software}
  {Engineering}}, {ICSE} '11, (New York, NY, USA), pp.~561--570, Association
  for Computing Machinery, May 2011.

\bibitem{zhao_image_2006}
F.~Zhao, Q.~Huang, and W.~Gao, ``Image {Matching} by {Normalized}
  {Cross}-{Correlation},'' in {\em 2006 {IEEE} {International} {Conference} on
  {Acoustics} {Speech} and {Signal} {Processing} {Proceedings}}, vol.~2,
  pp.~II--II, May 2006.

\bibitem{imagemagick_compare}
``{ImageMagick} {Comparison}.''
\newblock https://imagemagick.org/Usage/compare/.

\bibitem{ismael_castell-uroz_online_2020}
``{ORM},'' June 2020.
\newblock https://github.com/CBA-UPC/ORM.

\end{thebibliography}
\BgThispage

\end{document}